\documentclass{bmvc2k}
\usepackage{graphicx}
\usepackage{amsmath,amssymb,amsfonts}
\usepackage[binary-units=true]{siunitx}
\usepackage{xcolor}
\usepackage{multirow}

\usepackage{tikz}
\usetikzlibrary{shapes.arrows}
\usetikzlibrary{arrows.meta, arrows}
\usepackage{pgfplots}
\pgfplotsset{compat=newest}

\definecolor{tud0d}{cmyk/RGB/HTML}{0,0,0,.8/83,83,83/535353}
\definecolor{tud0c}{cmyk/RGB/HTML}{0,0,0,.6/137,137,137/898989}
\definecolor{tud0b}{cmyk/RGB/HTML}{0,0,0,.4/181,181,181/B5B5B5}
\definecolor{tud0a}{cmyk/RGB/HTML}{0,0,0,.2/220,220,220/DCDCDC}
\definecolor{tud1a}{cmyk/RGB/HTML}{.7,.4,0,0/93,133,195/5D85C3}
\definecolor{tud2a}{cmyk/RGB/HTML}{0.8,.2,0,0/0,156,218/009CDA}
\definecolor{tud3a}{cmyk/RGB/HTML}{0.7,0,.5,0/80,182,149/50B695}
\definecolor{tud4a}{cmyk/RGB/HTML}{.4,0,.8,0/175,204,80/AFCC50}
\definecolor{tud5a}{cmyk/RGB/HTML}{.2,0,.8,0/221,223,72/DDDF48}
\definecolor{tud6a}{cmyk/RGB/HTML}{0,.1,.7,0/255,224,92/FFE05C}
\definecolor{tud7a}{cmyk/RGB/HTML}{0,.3,.8,0/248,186,60/F8BA3C}
\definecolor{tud8a}{cmyk/RGB/HTML}{0,.6,.8,0 /238,122,52/EE7A34}
\definecolor{tud9a}{cmyk/RGB/HTML}{0,.8,.7,0/233,80,62/E9503E}
\definecolor{tud10a}{cmyk/RGB/HTML}{.2,.9,0,0/201,48,142/C9308E}
\definecolor{tud11a}{cmyk/RGB/HTML}{.6,.8,0,0/128,69,151/804597}
\definecolor{tud1b}{cmyk/RGB/HTML}{1,.6,0,0/0,90,169/005AA9}
\definecolor{tud2b}{cmyk/RGB/HTML}{1,.3,0,0/0,131,204/0083CC}
\definecolor{tud3b}{cmyk/RGB/HTML}{1,0,.6,0/0,157,129/009D81}
\definecolor{tud4b}{cmyk/RGB/HTML}{.5,0,1,0/153,192,0/99C000}
\definecolor{tud5b}{cmyk/RGB/HTML}{.3,0,1,0/201,212,0/C9D400}
\definecolor{tud6b}{cmyk/RGB/HTML}{0,.2,1,0/253,202,0/FDCA00}
\definecolor{tud7b}{cmyk/RGB/HTML}{0,.4,1,0/245,163,0/F5A300}
\definecolor{tud8b}{cmyk/RGB/HTML}{0,.7,1,0/236,101,0/EC6500}
\definecolor{tud9b}{cmyk/RGB/HTML}{0,1,.9,0/230,0,26/E6001A}
\definecolor{tud10b}{cmyk/RGB/HTML}{.4,1,0,0/166,0,132/A60084}
\definecolor{tud11b}{cmyk/RGB/HTML}{.7,1,0,0/114,16,133/721085}
\definecolor{tud1c}{cmyk/RGB/HTML}{1,.7,.2,0/0,78,138/004E8A}
\definecolor{tud2c}{cmyk/RGB/HTML}{1,.5,.2,0/0,104,157/00689D}
\definecolor{tud3c}{cmyk/RGB/HTML}{1,.2,.6,0/0,136,119/008877}
\definecolor{tud4c}{cmyk/RGB/HTML}{.6,.1,1,0/127,171,22/7FAB16}
\definecolor{tud5c}{cmyk/RGB/HTML}{.4,.1,1,0/177,189,0/B1BD00}
\definecolor{tud6c}{cmyk/RGB/HTML}{.2,.3,1,0/215,172,0/D7AC00}
\definecolor{tud7c}{cmyk/RGB/HTML}{.2,.5,1,0/210,135,0/D28700}
\definecolor{tud8c}{cmyk/RGB/HTML}{.2,.8,1,0/204,76,3/CC4C03}
\definecolor{tud9c}{cmyk/RGB/HTML}{.3,1,.9,0/185,15,34/B90F22}
\definecolor{tud10c}{cmyk/RGB/HTML}{.5,1,.3,0/149,17,105/951169}
\definecolor{tud11c}{cmyk/RGB/HTML}{.8,1,.2,0/97,28,115/611C73}
\definecolor{tud1d}{cmyk/RGB/HTML}{1,.9,.3,0/36,53,114/243572}
\definecolor{tud2d}{cmyk/RGB/HTML}{1,.7,.4,0/0,78,115/004E73}
\definecolor{tud3d}{cmyk/RGB/HTML}{1,.4,.7,0/0,113,94/00715E}
\definecolor{tud4d}{cmyk/RGB/HTML}{.7,.3,1,0/106,139,55/6A8B22}
\definecolor{tud5d}{cmyk/RGB/HTML}{.5,.2,1,0/153,166,4/99A604}
\definecolor{tud6d}{cmyk/RGB/HTML}{.4,.4,1,0/174,142,0/AE8E00}
\definecolor{tud7d}{cmyk/RGB/HTML}{.3,.6,1,0/190,111,0/BE6F00}
\definecolor{tud8d}{cmyk/RGB/HTML}{.4,.8,1,0/169,73,19/A94913}
\definecolor{tud9d}{cmyk/RGB/HTML}{.5,1,.9,0/156,28,38/961C26}
\definecolor{tud10d}{cmyk/RGB/HTML}{.7,1,.5,0/115,32,84/732054}
\definecolor{tud11d}{cmyk/RGB/HTML}{.9,1,.3,0/76,34,106/4C226A}

\newcommand \colorindicator[2]{%
	#1 {\textcolor{#2}{$\blacksquare\!\!\!\!\blacksquare$}}%
}

\title{OSS-Net: Memory Efficient High Resolution Semantic Segmentation of 3D Medical Data}

\addauthor{Christoph Reich}{christoph.reich@bcs.tu-darmstadt.de}{1}
\addauthor{Tim Prangemeier}{tim.prangemeier@bcs.tu-darmstadt.de}{1}
\addauthor{Özdemir Cetin}{oezdemir.cetin@bcs.tu-darmstadt.de}{1}
\addauthor{Heinz Koeppl}{heinz.koeppl@bcs.tu-darmstadt.de}{1}

\addinstitution{
	Centre for Synthetic Biology,\\
	Department of Electrical Engineering and Information Technology, Department of Biology,\\
	Technische Universität Darmstadt
}

\runninghead{Reich, Prangemeier, Cetin, Koeppl}{OSS-Net}

\def\etal{\emph{et al}\bmvaOneDot}

\begin{document}
	\maketitle
	
	\begin{abstract}
		Convolutional neural networks (CNNs) are the current state-of-the-art meta-algorithm for volumetric segmentation of medical data, for example, to localize COVID-19 infected tissue on computer tomography scans or the detection of tumour volumes in magnetic resonance imaging. A key limitation of 3D CNNs on voxelised data is that the memory consumption grows cubically with the training data resolution. Occupancy networks (O-Nets) are an alternative for which the data is represented continuously in a function space and 3D shapes are learned as a continuous decision boundary. While O-Nets are significantly more memory efficient than 3D CNNs, they are limited to simple shapes, are relatively slow at inference, and have not yet been adapted for 3D semantic segmentation of medical data. Here, we propose Occupancy Networks for Semantic Segmentation (OSS-Nets) to accurately and memory-efficiently segment 3D medical data. We build upon the original O-Net with modifications for increased expressiveness leading to improved segmentation performance comparable to 3D CNNs, as well as modifications for faster inference. We leverage local observations to represent complex shapes and prior encoder predictions to expedite inference. We showcase OSS-Net's performance on 3D brain tumour and liver segmentation against a function space baseline (O-Net), a performance baseline (3D residual U-Net), and an efficiency baseline (2D residual U-Net). OSS-Net yields segmentation results similar to the performance baseline and superior to the function space and efficiency baselines. In terms of memory efficiency, OSS-Net consumes comparable amounts of memory as the function space baseline, somewhat more memory than the efficiency baseline and significantly less than the performance baseline. As such, OSS-Net enables memory-efficient and accurate 3D semantic segmentation that can scale to high resolutions.\\[2pt]\begin{minipage}{0.75\textwidth}{\hspace{10pt} Code and trained models are available at \url{https://github.com/ChristophReich1996/OSS-Net}.}\end{minipage}
	\end{abstract}
	
	\section{Introduction} \label{sec:introduction}
Transferring established 2D deep learning solutions, such as CNNs, to dense 3D data entails various issues. A significant limitation of this approach, is the increased computational complexity and memory consumption for processing dense  3D volumes. Nonetheless, state-of-the-art approaches for segmenting 3D medical data rely on 3D CNNs \cite{cicek2016, milletari2016, myronenko2018, brugger2019, ji2020, chen2021}. The main drawback of training 3D CNNs, to learn a voxelised mapping, is the cubic growth of the memory and computational complexity with the training voxel resolution \cite{mescheder2019}. In practice, the available GPU memory limits the voxel resolution, resulting in trade-offs between memory usage, segmentation resolution, model capacity, and inference speed \cite{kamnitsas2016, myronenko2018, zhou2019}.\\
\indent Deep learning approaches that leverage 3D representations beyond dense voxelized volumes are available to overcome the cubical complexity of 3D CNNs \cite{mescheder2019, qi2017b, wang2018, gkioxari2019}. Implicit neural representations, such as occupancy networks \cite{mescheder2019}, have been proposed for various 3D learning tasks \cite{chen2019, park2019, michalkiewicz2019}. O-Nets learn a continuous decision boundary in a function space to represent 3D shapes. Predicting occupancy values in training, instead of a dense voxelised representation, enables O-Nets to be significantly more memory efficient than CNNs on 3D data \cite{mescheder2019}. Despite the memory efficiency, O-Nets are relatively slow at inference and their expressiveness is limited in comparison to CNNs \cite{mescheder2019, peng2020}. To the best of the author's knowledge, O-Nets have not yet been modified to jointly overcome these shortcomings in 3D segmentation, or to segment 3D medical data.\\
\begin{figure}[!ht]
	\centering
	\vspace{-0.5cm}
	\includegraphics[width=0.25\columnwidth, angle=-90,origin=c, clip, trim=25px 25px 25px 25px]{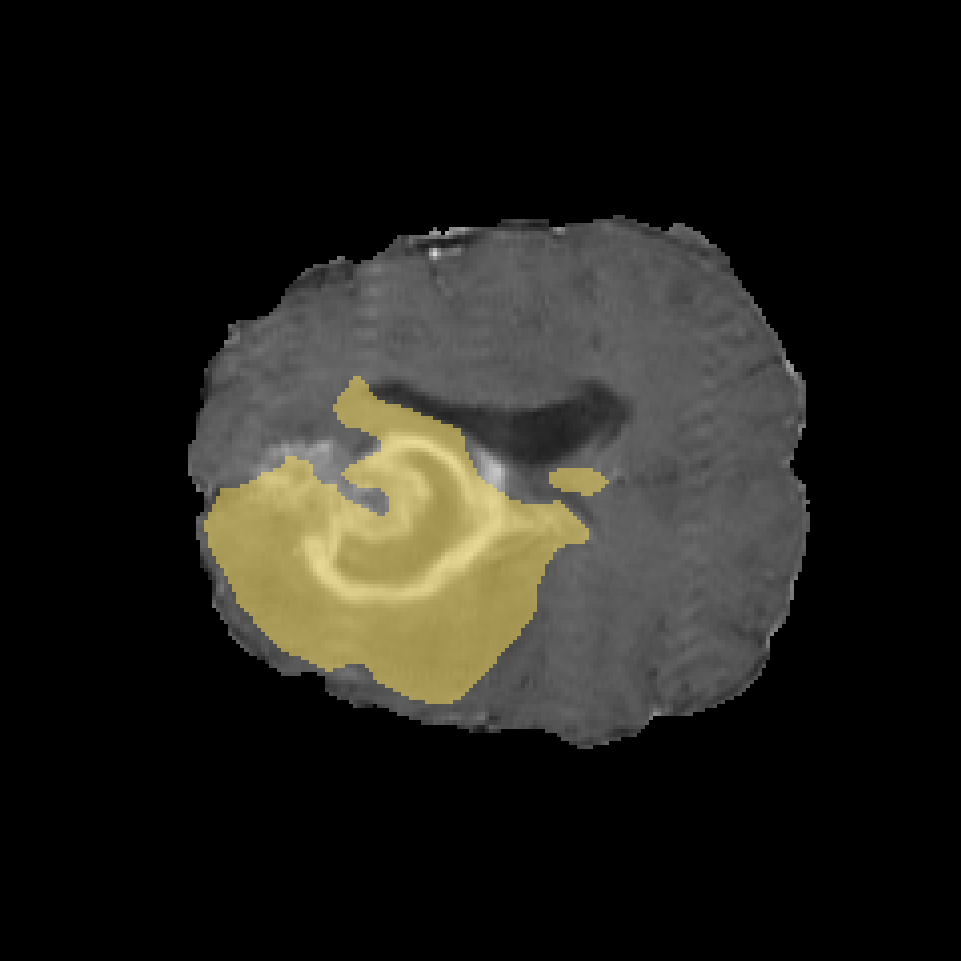}\includegraphics[clip, trim=400px 100px 300px 200px, width=0.25\columnwidth]{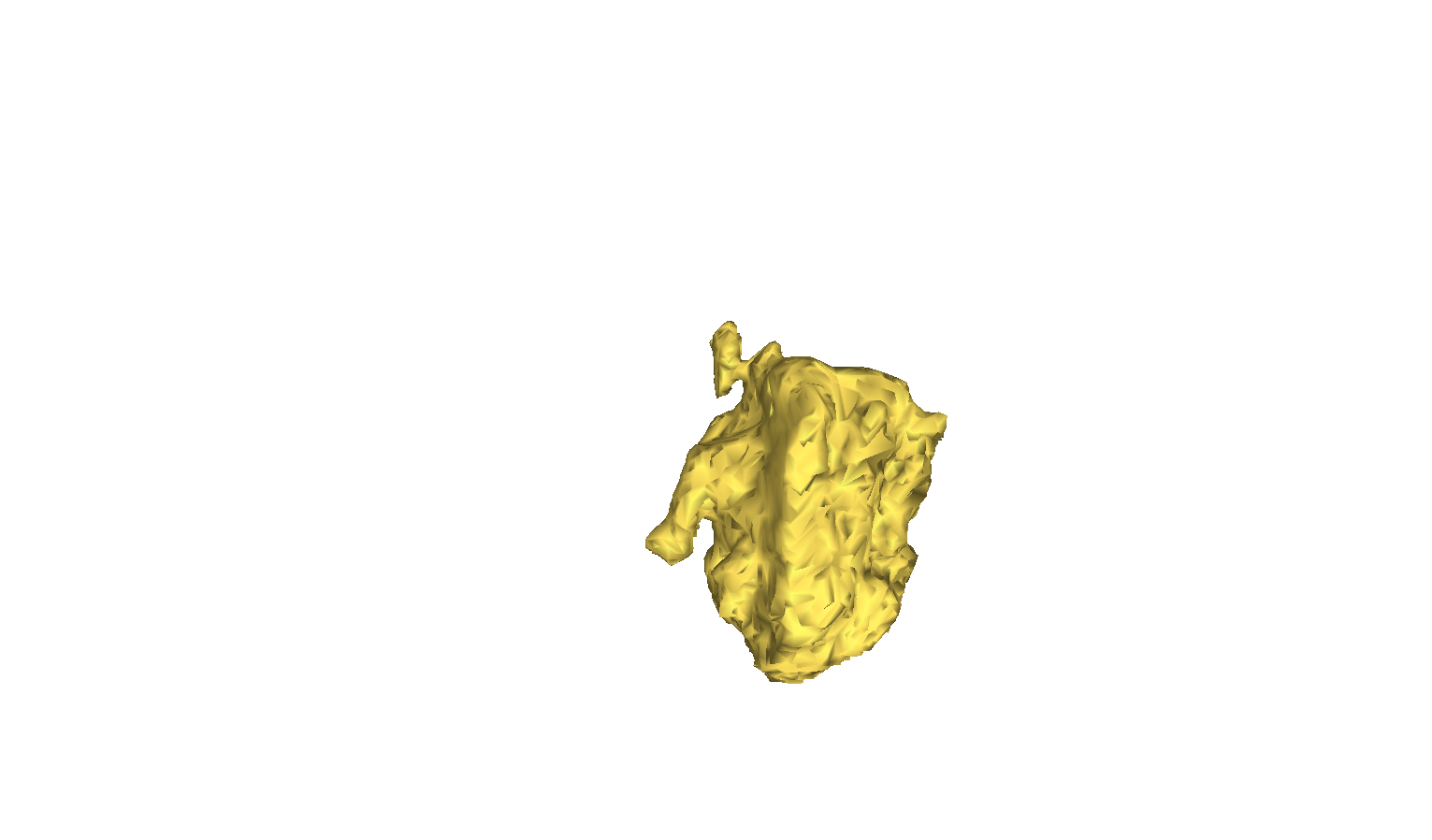}\includegraphics[width=0.25\columnwidth, angle=-90,origin=c, clip, trim=25px 25px 25px 25px]{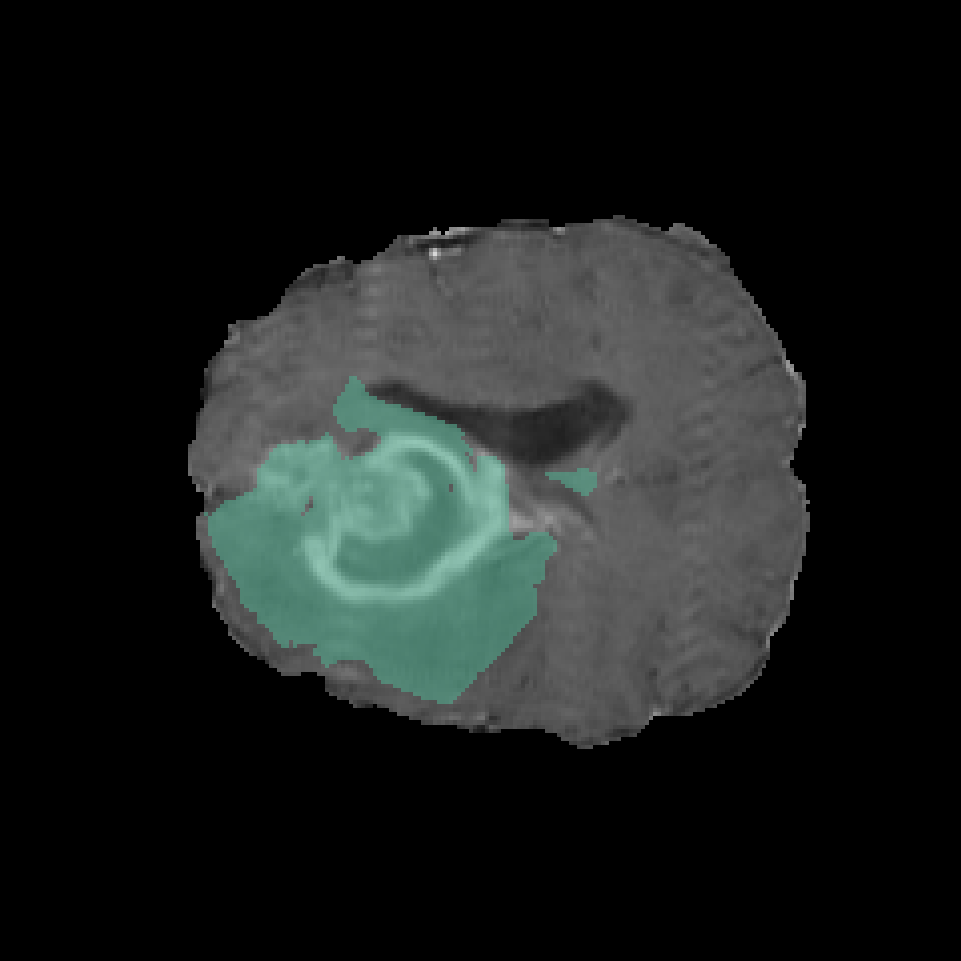}\includegraphics[clip, trim=400px 100px 300px 200px, width=0.25\columnwidth]{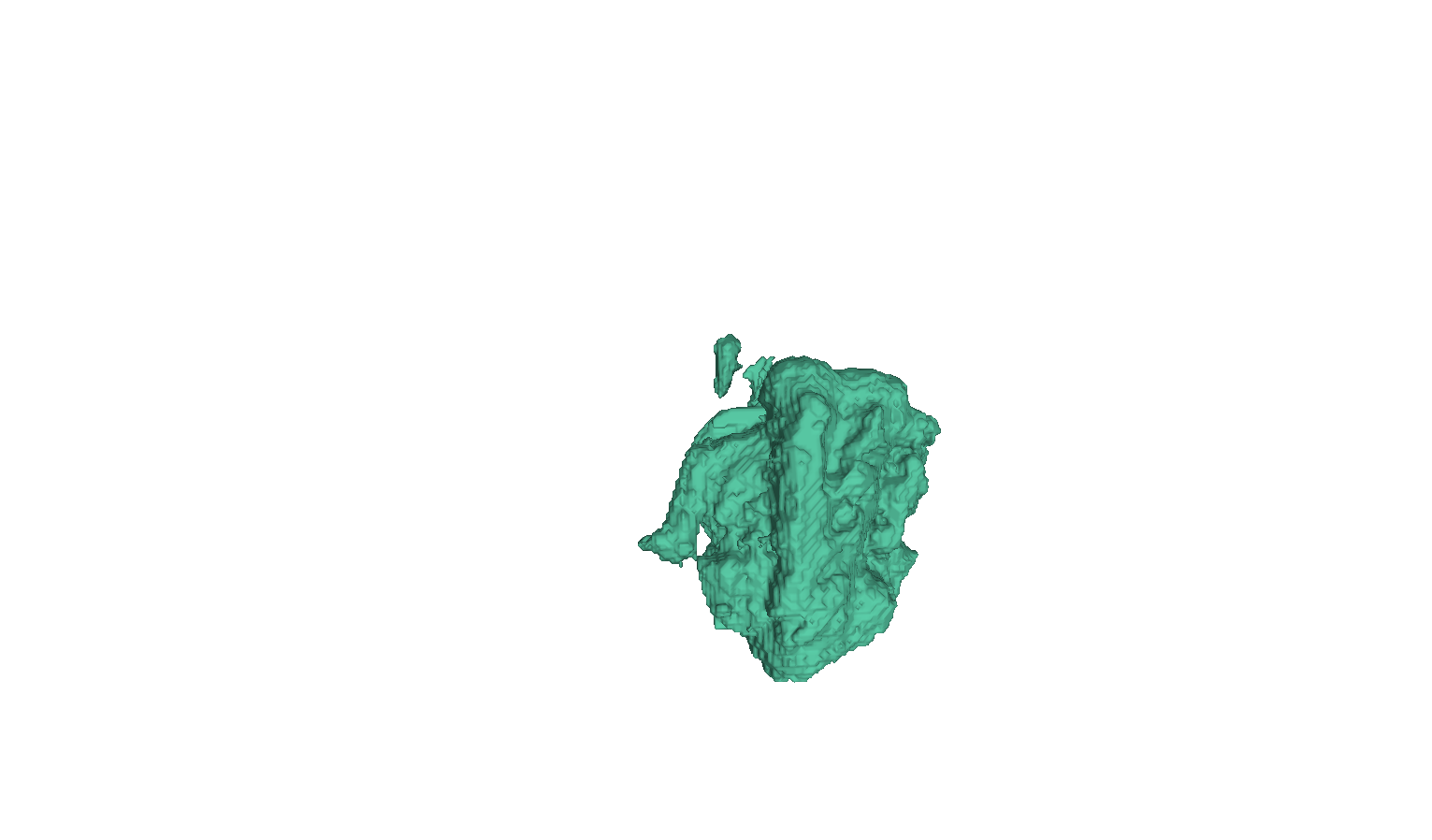}\\
	\caption{Brain tumour segmentation results of OSS-Net (config. C) on the BraTS 2020 dataset \cite{bakas2018}. Brain tumour prediction in \colorindicator{yellow}{tud6a!50} and label in \colorindicator{green}{tud3a}. 2D MRI slice (Tc1 modality) overlaid with the corresponding voxelized prediction or label on the left and the corresponding extracted mesh on the right. Best viewed in color.}
	\label{fig:firstfig}
\end{figure}

In this study, we propose Occupancy Networks for Semantic Segmentation to combine the performance of 3D CNNs with the memory efficiency of O-Nets while avoiding the drawbacks of the respective methods. We modify the original O-Net architecture by leveraging local observations. This enables OSS-Net to represent more complex 3D shapes, such as the fine tumour structure depicted in Figure \ref{fig:firstfig}. Inference speed is improved by utilising a prior prediction of the OSS-Net encoder as an initial state of the inference approach. We demonstrate the performance on the BraTS 2020 dataset for 3D brain segmentation \cite{menze2014, bakas2017, bakas2018} and the LiTS dataset for 3D liver segmentation \cite{bilic2019}. On both datasets, OSS-Net significantly surpass the original O-Net's segmentation performance, while being similarly memory efficient. OSS-Net performs on par in segmentation accuracy with the voxelised baseline (3D residual U-Net \cite{cicek2016, myronenko2018}) for brain tumour segmentation while falling slightly short of this baseline on the smaller LiTS dataset. OSS-Net makes it feasible to scale to high voxel resolutions ($512^3$) while the memory consumption remains manageable.
	\section{Related Work} \label{sec:related_work}
\indent Recent approaches towards memory efficient processing of three-dimensional data with deep neural networks can be clustered into two categories. The first category is based around making existing 3D CNNs architectures more efficient. RevNet by Gomez \etal \cite{gomez2017} is an example of this and employs mathematically reversible building blocks, avoiding the need to store all activations during training. Brugger \etal employed this approach for semantic segmentation of 3D medical data, achieving a $30\%$ reduction in memory cost in comparison to a baseline 3D U-Net \cite{brugger2019}. Another avenue towards increased efficiency of existing CNNs is to utilise sparse convolutions, as proposed by \cite{graham2018, choy2019}. This approach is of limited applicability to medical data, where the required spare data is seldom given \cite{menze2014, bilic2019}. A practical approach for increasing the memory efficiency of 3D biomedical segmentation with CNNs is to reduce the 3D segmentation problem to a 2D problem by slicing and utilising a 2D CNN, for example a 2D U-Net \cite{zhou2016, zhou2017, perslev2019, prangemeier2020b, prangemeier2021}. While these approaches can achieve high memory efficiency, they inherit the strong limiting assumption that a certain 3D segmentation task can be solved in 2D \cite{gibson2018, perslev2019}, neglecting 3D relations and limiting their applicability to general 3D problems \cite{roth2018}. Recent refined slicing-based approaches also include an additional hand-designed fusion step, adding complexity to the segmentation pipeline \cite{zhou2016, perslev2019}.\\
\indent The second category is to consider alternative approaches to represent 3D data beyond voxelised volumes, such as 3D point clouds or 3D meshes. Both representations have recently been used to perform 3D semantic segmentation \cite{qi2017b, wang2018, gkioxari2019, guo2020}. 3D point cloud methods typically require heavy post-processing, and the available GPU memory limits 3D mesh approaches \cite{mescheder2019}. Furthermore, both 3D point cloud and 3D mesh approaches are challenging to apply to dense and high-resolution 3D medical data.\\
\indent An alternative representation that is potentially significantly more efficient was recently proposed by Mescheder \etal \cite{mescheder2019}. The proposed O-Net learns a continuous decision boundary in a 3D function space to reconstruct a 3D shape. The resulting continuous decision boundary can be extracted at an arbitrary resolution. Compared to voxelised architectures, O-Nets have a vastly higher memory efficiency since the typically used CNN decoder is replaced with a fully-connected occupancy decoder \cite{mescheder2019}.\\
\indent Occupancy networks and related implicit representation approaches, such as \cite{chen2019, park2019, michalkiewicz2019}, are memory efficient yet limited to represent simple 3D shapes and do not generalize well to unseen shapes \cite{peng2020}. This issue is mainly caused by the lack of local features in the occupancy decoder \cite{peng2020}. To overcome this limitation, multiple extensions have been proposed \cite{xu2019,peng2020, genova2020, saito2020, chibane2020, mihajlovic2021}. These extensions are either not applicable to the task of 3D semantic segmentation, require highly complex architectures, or sacrifice memory efficiency.

	\section{Occupancy Networks for Semantic Segmentation}
\label{sec:oss-net}
An occupancy network is described by the learnable mapping $f_{\theta}:\mathbb{R}^3\times\mathcal{X}\to\left[0, 1\right]$ \cite{mescheder2019}. The mapping is parameterised by $\theta$. The inputs are a 3D location $p\in\mathbb{R}^3$ and an observation $x\in\mathcal{X}$. The output is the occupancy probability $o\in\left[0, 1\right]$ that describes whether the input location lies within or outside a continuous decision boundary. In the case of semantic segmentation, the observation $x$ is a 3D volume that is encoded by a 3D CNN to a global latent representation $x_{l}$. Since only a global representation $x_{l}$ of the observation $x$ is produced, the occupancy network $f_{\theta}\left(p, x\right)$ cannot capture small details in practice \cite{peng2020}.\\ 
\indent To overcome the lack of local information in our OSS-Net occupancy encoder, we extend the original learnable mapping $f_{\theta}$ to
\begin{equation}\label{eq:ossnet}
    f_{\theta}:\mathbb{R}^3\times\mathcal{X}\times\mathcal{Z}\to\left[0, 1\right],
\end{equation}
with a local observation $z\in\mathcal{Z}$ as an additional input. The local observation $z$ is a local 3D patch sampled from the global observation $x$ centered at the 3D location $p$ which is encoded to a local latent representation. Employing a local latent representation for each location as the input to the occupancy encoder solves the issue of missing local information. Equation \ref{eq:ossnet} can readily be adapted to multi-class segmentation by chaining the last layer to predict a softmax vector instead of a scalar occupancy probability.
\subsection{Architecture}
\label{subsec:architecture}
\begin{figure}[!ht]
	\centering
	\includegraphics[scale=1.0]{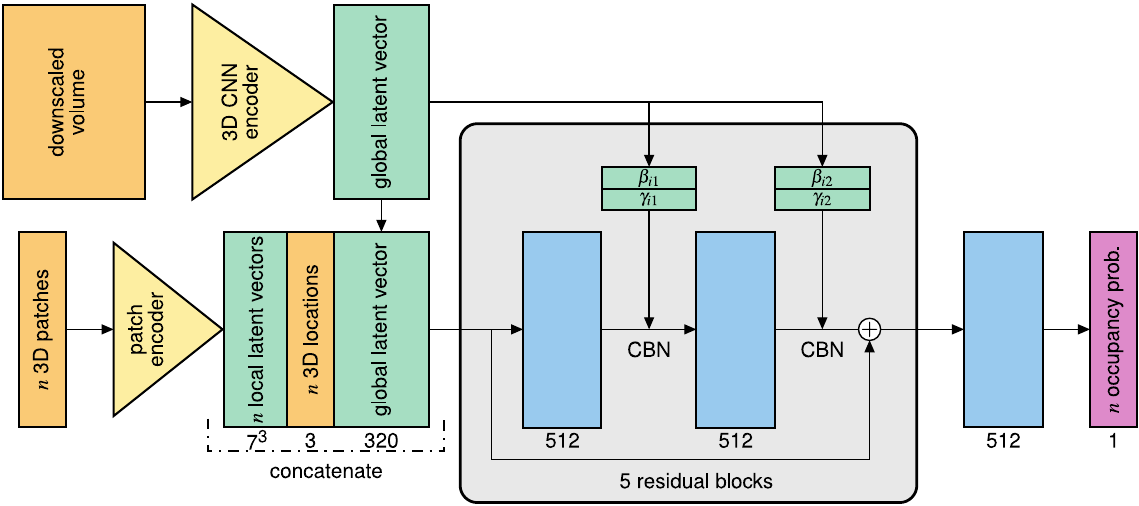}
	\caption{Architecture of the OSS-Net, with downscaled global volume, $n$ local 3D patches, and $n$ 3D locations (in \colorindicator{orange}{tud7b!50}) as inputs. The 3D CNN encoder (in \colorindicator{yellow}{tud6a!50}) extracts a global latent vector (in \colorindicator{green}{tud3a!50}). The patch encoder produces $n$ local latent vectors (in \colorindicator{green}{tud3a!50}). The global and local latent vectors as well as the $n$ 3D locations are concatenated and fed into five residual fully connected blocks with conditional batch normalization ($\beta$ \& $\gamma$ predicted parameters) to produce $n$ occupancy probability predictions (in \colorindicator{purple}{tud10a!50}). Best viewed in color.}
	\label{fig:oss-net-architecture}
\end{figure}

We implemented OSS-Net, described by Equation \ref{eq:ossnet} as a deep neural network composed of three main components. An overview of the OSS-Net architecture is provided in Figure \ref{fig:oss-net-architecture}. First, the 3D CNN encoder maps a downscaled input volume to a global latent vector. Second, the patch encoder, consisting of two 3D convolutions, encodes $n$ 3D patches and produces $n$ local latent vectors. Third, based on the global and the $n$ corresponding local latent vectors, the fully connected occupancy decoder produces an occupancy probability prediction for each of the $n$ input coordinates.\\

\begin{figure}[!ht]
	\centering
	\includegraphics[scale=1.0]{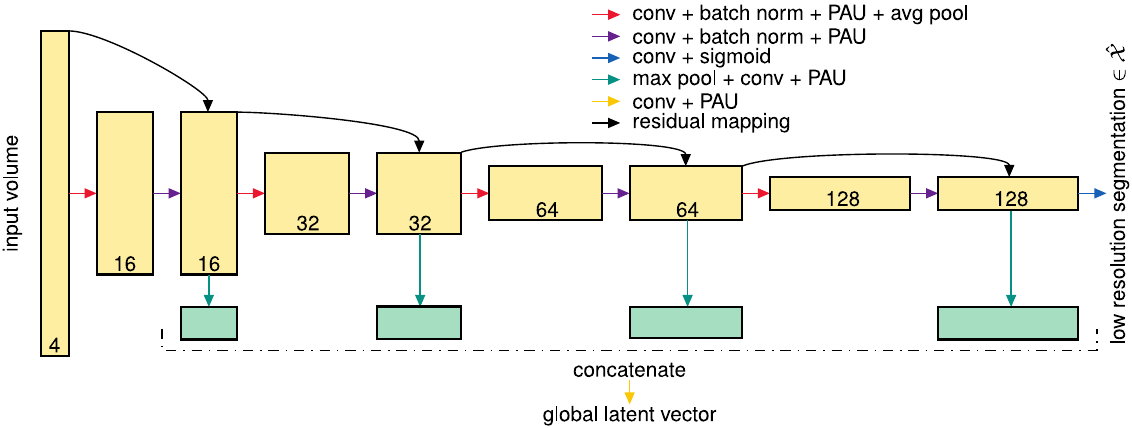}
	\caption{Residual 3D CNN encoder with output skip-connections. Operations indicated with an arrow, feature tensors colored in \colorindicator{yellow}{tud6a!50} and intermediated latent tensors visualized in \colorindicator{green}{tud3a!50}. Channel dimensions shown in feature tensors are corresponding to BraTS network setting. Best viewed in color.}
	\label{fig:oss-net-encoder}
\end{figure}

\indent The task-specific 3D CNN encoder (Fig. \ref{fig:oss-net-encoder}) exhibits a ResNet \cite{he2016} like architecture. We utilise output skip-connections in the encoder, as introduced by \cite{xu2019}. These incorporate features of each encoder stage directly into the global latent vector. The output of the lowest encoder stage predicts a low-resolution voxelized segmentation of the input volume. This segmentation is used in an auxiliary loss during training (Sec. \ref{subsec:training}). At inference the segmentation is used to speed up the dense segmentation extraction approach (Sec. \ref{subsec:inference}).\\

\subsection{Training and Evaluation}
\label{subsec:training}
Training the neural network, $f_{\theta}\left(p, x, z\right)$ (Eq. \ref{eq:ossnet}), was approached by sampling $n$ 3D locations $p$ and corresponding local patches $z$ from the input volume $x$ to be segmented. The predicted occupancy probabilities were supervised by a binary cross-entropy loss (first term)
\begin{multline}\label{eq:loss}
    \mathcal{L}=-\frac{1}{k\,n}\sum_{i=1}^{k}\sum_{j=1}^{n}\left[{o}_{ij}\log(f_{\theta}\left(p_{ij}, x_{i}, z_{ij}\right)) + \left(1-{o}_{ij}\right)\log(1-f_{\theta}\left(p_{ij}, x_{i}, z_{ij}\right))\right]\\
    -\alpha\,\frac{1}{k\,w}\sum_{i=1}^{k}\sum_{m=1}^{w}\left[y_{im}\log(f_{\theta}^{e}\left(x_{i}\right)_{m})) + (1-y_{im})\log(1 - f_{\theta}^{e}\left(x_{i}\right)_{m})\right].
\end{multline}
The main loss (first term) is computed over $n$ sampled locations, with it's corresponding ground truth label ${o}$, in a mini-batch of size $k$.\\
\indent Sampling $n$ 3D locations from the global volume is performed by uniform sampling and border region sampling. In the border region sampling, one half of the locations are sampled uniformly from the $25$ voxel wide border of the ground truth segmentation hull. The other half of the location is sampled uniformly over the whole volume.\\
\indent To guide the encoder to learn relevant features for the segmentation, we employed and auxiliary endocer loss in addition the main occupancy loss (Equation \ref{eq:loss} second term). The auxiliary loss (Equation \ref{eq:loss} second term) is also a binary cross-entropy loss, computed between the low-resolution voxelized segmentation (voxels indexed by $m$) prediction of the encoder segmentation mapping $f_{\theta}^e:\mathcal{X}\to\hat{\mathcal{X}}$ and the downsampled ground truth voxelised label $y$. The weighting factor $\alpha\in\mathbb{R}$ was $0.1$.\\
\indent We followed the original O-Net approach, uniformly sampling $2^{17}$ locations and patches from the global volume, to validate the OSS-Net during training \cite{mescheder2019}. The resulting occupancy predictions and corresponding labels were utilised to compute the Intersection over Union (IoU) $\frac{\left|\mathbf{o}\,\cap\,\hat{\mathbf{o}}\right|}{\left|\mathbf{o}\,\cup\,\hat{\mathbf{o}}\right|}$ and the Dice score $\frac{2\left|\mathbf{o}\,\cap\,\hat{\mathbf{o}}\right|}{\left|\mathbf{o}\right| + \left|\hat{\mathbf{o}}\right|}$ between the sampled ground truth labels $\mathbf{o}$ and the corresponding predictions $\hat{\mathbf{o}}$.
\subsection{Inference} \label{subsec:inference}
At inference time, the goal is to extract the predicted decision boundary of the OSS-Net. This can be done by utilizing the Multiresolution IsoSurface Extraction (MISE) \cite{mescheder2019} algorithm proposed with the original O-Net. We built upon this approach with the aim to reduce the runtime while preserving the prediction accuracy.\\
\begin{figure}[!ht]
    \centering
    \includegraphics[scale=1.0]{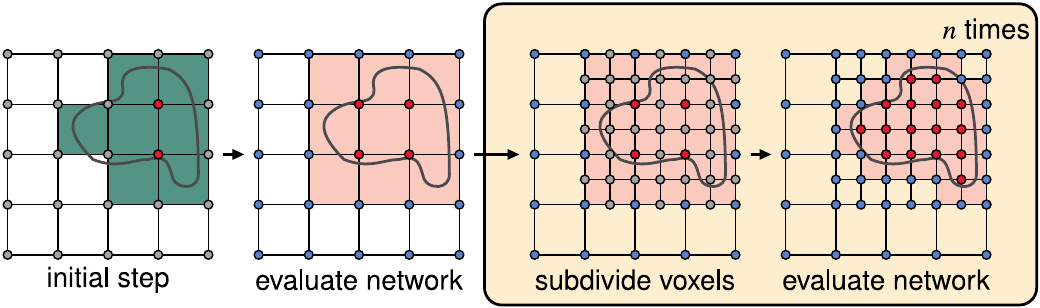}
    \caption{Our improved dense segmentation extraction approach in 2D. Initial upsampled and thresholded segmentation of the 3D CNN encoder in \colorindicator{green}{tud3d!70}, occupied coordinates in \colorindicator{red}{tud9b}, unoccupied coordinates in \colorindicator{blue}{tud1a}, coordinates to be evaluated in \colorindicator{gray}{tud0b}, and the current voxelized segmentation in \colorindicator{pink}{tud9b!20}. Best viewed in color.}
    \label{fig:inference}
\end{figure}
\indent MISE starts by building an octree and incrementally evaluates the continuous decision boundary \cite{mescheder2019}. In the next step, the produced dense prediction is used to extract a mesh which is subsequently smoothed \cite{mescheder2019}. An example of a mesh extracted with this approach is shown in Figure \ref{fig:firstfig}. Medical use-cases typically only require dense voxelised predictions and we omitted the mesh extraction and smoothing in practice. We refer to MISE without mesh extraction and smoothing as the original O-Net inference approach.\\
\indent Our modified inference approach (Fig. \ref{fig:inference}) utilises the low-resolution dense prediction of the encoder as the initial state of the octree. This results in faster inference since fewer locations have to be evaluated by the OSS-Net decoder. The low-resolution encoder segmentation is thresholded and used to determine which locations are occupied in the initial octree. All unoccupied locations are evaluated by the OSS-Net decoder to complete the first evaluation stage. In the following stages, the voxels of the current segmentation are subdivided and then refined by additional network evaluations. This process is repeated until the desired resolution is reached. If the desired resolution surpasses the resolution of the data sample, patches can be queried from the sample using trilinear interpolation. A 2D visualization of OSS-Net's expedited inference approach is presented in Figure \ref{fig:inference}.\\
	\section{Experiments}
\label{sec:experiments}
We analyse the segmentation performance (Tab. \ref{tab:segmentationresults} \& \ref{tab:samplingresults}), memory consumption (Tab. \ref{tab:memoryresults} \& Fig. \ref{fig:memory}), and inference speed (Fig. \ref{fig:runtime}) of the proposed OSS-Net in binary semantic segmentation experiments on the BraTS 2020 and the LiTS challenge dataset, and compare these to an original O-Net and a voxelised CNN baseline. We investigate the efficacy of the proposed modifications by adding these to the original O-Net incrementally. All segmentation results are based on 3 training runs with different random seeds, with the best result reported.
\subsection{Baselines and Implementation Details}
An original O-Net \cite{mescheder2019} serves as our function space baseline. We employed a basic 3D ResNet encoder without output skip-connection in the original O-Net. Five residual fully connected blocks with 512 features were utilised as the occupancy encoder. Conditional batch normalization \cite{de2017} between the encoder and decoder was used alongside a concatenation of the global latent vector with the input of the occupancy encoder as in \cite{niemeyer2019} (Fig. \ref{fig:oss-net-architecture}).\\
\indent All function space models were trained for 50 epochs with a combination of the Lookahead  \cite{zhang2019} and RAdam \cite{liu2019} optimiser. The learning rate was  $3\times 10^{-4}$ except for the weights of the Pad\'{e} Activation Units \cite{molina2020}, where a learning rate of $10^{-2}$ was employed. Each learning rate was decreased by an order of magnitude after 20 and 30 epochs. The Lookahead optimiser parameters $k$ and $\alpha$ were $5$ and $0.8$, respectively. \\
\indent A 3D residual U-Net \cite{cicek2016, milletari2016, zhang2018, myronenko2018} trained on the binary cross-entropy loss served as our voxelised performance baseline. The U-Net encoder is comprised of four residual blocks with 8, 32, 128, and 384 filters, respectively. In the decoder three residual blocks with 128, 32, and 8 respectively are utilised. Group normalization was employed in both the encoder and decoder \cite{wu2018}. The voxelised baseline was trained for 20 epochs on the same optimiser configuration as the function space approaches with a learning rate of $10^{-3}$.\\
\indent We employed a 2D residual U-Net \cite{zhang2018} with group normalization \cite{wu2018} as a memory-efficient slicing baseline. Five residual encoder blocks (128, 256, 384, 512, and 640 filters) and four residual encoder blocks (512, 384, 256, and 128 filters) were used. We used the same loss and optimiser configuration (learning rate $10^{-4}$) as for the voxelised baseline for training (50 epochs). On each dataset, we trained three networks for three separate slice orientations, to investigate and demonstrate their effect on segmentation performance. We slice along each spatial dimension (height, width, depth) and choose the best performing orientation, for a fair comparison.\\
\indent All training runs were performed on two Nvidia V100 $16\si{\giga\byte}$. For all function space models, we employed a batch size of 8 on the BraTS 2020 dataset experiments and a batch size of 4 on the LiTS dataset. The voxelised baseline was trained with a batch size of 2. The slicing baseline used batch size of 32 and 12 for the BraTS 2020 and LiTS datasets, respectively. All inference tests were performed on a single consumer-grade Nvidia RTX 2080 Ti.\\
\subsection{Datasets and Data Augmentation}
The BraTS 2020 \cite{menze2014, bakas2017, bakas2018} and the LiTS \cite{bilic2019} dataset were utilised for binary semantic segmentation. We trained the networks for segmenting whole brain tumours and full livers on the the BraTS 2020 and LiTS datasets, respectively. The BraTS 2020 dataset was utilised in its native resolution of $240 \times 240 \times 155$, which is zero-padded to $256 \times 256 \times 160$. The variably sized CT scans and pixel-wise labels of the LiTS dataset were resized to a fixed voxel resolution of $512^3$. The voxelised baseline was trained on downscaled CT scans with $224^3$ voxels, to match the limited available GPU memory of $16\si{\giga\byte}$. At inference time the predictions are trilinearly upsampled to the native training resolution. For reproducibility we only employed the publicly available samples of the BraTS 2020 and LiTS dataset. We split the BraTS 2020 dataset randomly into $320$ training and $45$ validation samples, as well as the LiTS dataset into $111$ training and $20$ validation samples. Flipping, brightness adjustment, and Gaussian noise injection was used for data augmentation. Each augmentation was applied with a probability of $0.5$ for all training runs.
\subsection{Experimental Results}
\begin{figure}[!ht]
    \centering
    \begin{minipage}[t]{.485\textwidth}
        \centering
        \includegraphics[scale=1.0]{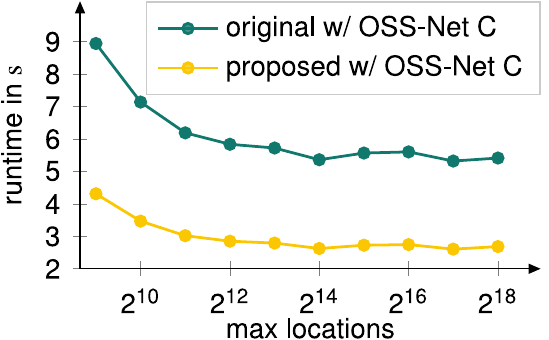}
        \caption{Inference runtimes for different maximum locations per encoder forward pass on the BraTS 2020 validation set. Standard O-Net inference approach in \colorindicator{green}{tud3d}, our improved inference approach in \colorindicator{yellow}{tud6b}. OSS-Net variant C employed.}
        \label{fig:runtime}
    \end{minipage}\hfill%
    \begin{minipage}[t]{.485\textwidth}
        \centering
        \includegraphics[scale=1.0]{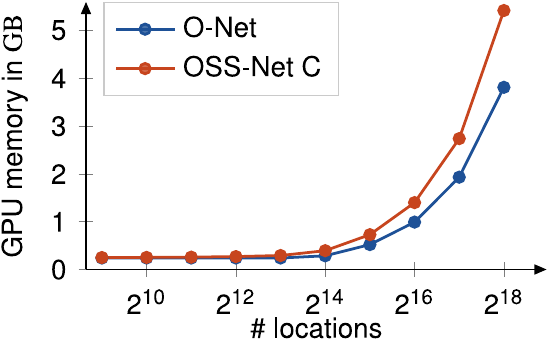}
        \caption{GPU memory consumption of a full inference forward pass for different number of sampled locations and patches (only OSS-Net C). O-Net baseline in \colorindicator{blue}{tud1c}, our OSS-Net C in \colorindicator{orange}{tud8c}. BraTS 2020 resolution utilised.}
        \label{fig:memory}
    \end{minipage}
\end{figure}

We observed a high impact of the sampling strategy on the segmentation performance, in preliminary experiments. Therefore, we conducted experiments to investigate the effect of different sampling strategies and the number of sampled locations, summarised in Table \ref{tab:samplingresults}.\\
\begin{table}[th!]
	\caption{Numerical comparison of different sampling strategies on the BraTS 2020 and LiTS dataset. OSS-Net variant C (Tab. \ref{tab:segmentationresults}) utilised.}
	\setlength\extrarowheight{0.605pt}
	\centering
	\label{tab:samplingresults}
	\begin{footnotesize}
	    \begin{tabular}{@{\extracolsep{4pt}}>{\raggedright\arraybackslash}p{0.2375\columnwidth}>{\centering\arraybackslash}p{0.2\columnwidth}>{\centering\arraybackslash}p{0.08\columnwidth}>{\centering\arraybackslash}p{0.08\columnwidth}>{\centering\arraybackslash}p{0.08\columnwidth}>{\centering\arraybackslash}p{0.08\columnwidth}}
	\hline
	 &  & \multicolumn{2}{c}{BraTS 2020}  & \multicolumn{2}{c}{LiTS} \\ 
	\cline{3-4} \cline{5-6}
	Sampling strategy & Samples & Dice $\uparrow$ & IoU $\uparrow$ & Dice $\uparrow$ & IoU $\uparrow$ \\ 
	\hline
	Uniform & $2^{13}$ & 0.8816 & 0.7941 & 0.6844 & 0.5248 \\
	Uniform & $2^{14}$ & \textbf{0.8842} & \textbf{0.7991} & 0.6317 & 0.4709 \\
	Uniform & $2^{15}$ & 0.8820 & 0.7944 & 0.6997 & 0.5427 \\
	\hline
	Border & $2^{13}$ & 0.8380 & 0.7367 & 0.7596 & 0.6168 \\
	Border & $2^{14}$ & 0.8353 & 0.7333 & \textbf{0.7616} & \textbf{0.6201} \\
	Border & $2^{15}$ & 0.8341 & 0.7312 & 0.7559 & 0.6105 \\
	\hline
\end{tabular}
	\end{footnotesize}
\end{table}
\indent On the BraTS 2020 dataset, the uniform sampling outperforms the border sampling approach, while results on the LiTS dataset are in favor of the border sampling approach. The best segmentation results were achieved for $2^{14}$ sampled location, regardless of the dataset. The segmentation performence dropped slighlty for $2^{13}$ sampled locations, and merely a single training run converged to a competitive result. The worst LiTS training run with $2^{13}$ samples and border sampling converged to an IoU of $0.5365$. Based on these preliminary findings, we employed uniform sampling for the experiments on the BraTS 2020 dataset and border sampling for the LiTS dataset experiments, both with $2^{14}$ sampled locations.\\
\indent The segmentation results achieved for various configurations of the proposed OSS-Net, the voxelised and slicing CNN baselines, as well as the function space baseline are summarised in Table \ref{tab:segmentationresults}. All proposed variants of the OSS-Net outperform the original O-Net in segmentation performance, on both the BraTS 2020 and the LiTS datasets. The OSS-Net variant A improves the IoU by $0.203$ over the original O-Net on the BraTS 2020 dataset. Employing the proposed encoder with output skip-connections and the auxiliary loss (variant C \& D) increases the Dice score further to $0.8842$ and $0.8774$, respectively. Using larger patches with a size of $14^3$ (variants B and D) lead to no significant change in performance. \\
\indent The best performing OSS-Net variant C performs on par to the voxelised baseline on the BraTS 2020 dataset, which was better on the smaller ($\sim 100$ training samples) LiTS dataset. In comparison to the slicing baseline, OSS-Net variant C achieves better segmentation performance on both datasets. The performance gap, between the explicit 3D methods (3D residual U-Net, OSS-Net variant C) and the slicing-baseline may be caused by the inherently limited spatial context of the 2D residual U-Net \cite{roth2018, gibson2018}. The slicing baseline is sensitive to slice orientation with significantly different performance for the different slice orientations (Tab. \ref{tab:samplingresults} footnote).\\
\begin{table}[!ht]
	\caption{Semantic segmentation results of our approaches and baselines on validation data.}
	\setlength\extrarowheight{0.605pt}
	\centering
	\label{tab:segmentationresults}
	\begin{footnotesize}
	    \begin{tabular}{@{\extracolsep{4pt}}>{\raggedright\arraybackslash}p{0.47\columnwidth}>{\centering\arraybackslash}p{0.085\columnwidth}>{\centering\arraybackslash}p{0.085\columnwidth}>{\centering\arraybackslash}p{0.085\columnwidth}>{\centering\arraybackslash}p{0.085\columnwidth}}
	\hline
	 & \multicolumn{2}{c}{BraTS 2020}  & \multicolumn{2}{c}{LiTS} \\ 
	\cline{2-3} \cline{4-5}
	Model/OSS-Net Configuration & Dice $\uparrow$ & IoU $\uparrow$ & Dice $\uparrow$ & IoU $\uparrow$ \\ 
	\hline
	3D residual U-Net (voxelised baseline) & \textit{0.8827} & \textbf{0.7995} & \textbf{0.7888} & \textbf{0.6558} \\
	2D residual U-Net (slicing baseline) & $\;\,$0.8589\textsuperscript{$\star$} & $\;\,$0.7658\textsuperscript{$\star$} & $\;\,$0.6674\textsuperscript{$\dagger$} & $\;\,$0.5233\textsuperscript{$\dagger$} \\
	O-Net (function space baseline) \cite{mescheder2019} & 0.7016 & 0.5615 & 0.6506 & 0.4842 \\ 
	\hline
	+ patch encoder w/ small patches $7^3$ (OSS-Net A) & 0.8592 & 0.7644 & 0.7127 & 0.5579 \\ 
	+ avg. pooled intermediate patches $14^3$ (OSS-Net B) & 0.8541 & 0.7572 &  0.7585 & 0.6154 \\ 
	A + encoder skip-con. \& aux. loss (OSS-Net C) & \textbf{0.8842} & \textit{0.7991} & \textit{0.7616} & \textit{0.6201} \\ 
	B + encoder skip-con. \& aux. loss (OSS-Net D) & 0.8774 & 0.7876 & 0.7566 & 0.6150 \\ 
	\hline
	\multicolumn{5}{l}{\textsuperscript{$\star$} Best slice depth orientated (height slice IoU 0.7269, Dice 0.8360; width slice IoU 0.7300, Dice 0.8369).}\\
	\multicolumn{5}{l}{\textsuperscript{$\dagger$} Best slice width orientated (height slice IoU 0.2732, Dice 0.1673; depth slice did not converge).}\\
\end{tabular}

	\end{footnotesize}
\end{table}
\indent The memory consumption is presented in Table \ref{tab:memoryresults}. OSS-Net C is 5 times more memory efficient than the voxelised baseline during training on the BraTS 2020 dataset, and 10 times more efficient for inference with $2^{14}$ locations. On the LiTS dataset, OSS-Net, operating at the native resolution ($512^3$), remains significantly more memory efficient than the voxelised baseline on a reduced resolution ($224^3$). In comparison to the function space baseline, O-Net, the proposed OSS-Net variant C requires a maximum of $10\%$ more memory at training.\\ 
\indent The slicing baseline requires less memory during training than the OSS-Nets. On the BraTS 2020 dataset, the 2D residual U-Net consumes half as much memory as the OSS-Net variant C at training. On the LiTS dataset with a resolution of $512^3$, the difference in training memory usage between the slicing baseline and OSS-Net variant C reduces to $0.9\si{\giga\byte}$. At inference, the memory for the OSS-Net variant C and the slicing baseline is roughly similar.\\
\indent Processing a single slice per forward pass of the 2D residual U-Net results in a runtime of $5\si{\second}$ for a full prediction on the BraTS 2020 dataset. When Utilizing multiple slices per forward pass this is reduced to $2.5\si{\second}$, similar to the inference runtime of the OSS-Net (characteristically $2.6\si{\second}$, see Fig. \ref{fig:runtime}). Processing multiple slices, per forward pass of the 2D U-Net, yields a linear increase in memory usage relative to the number of processed slices, resulting in a trade-off between inference speed and memory consumption, which OSS-Net avoids. \\
\begin{table}[!ht]
	\caption{GPU memory consumption of our networks and baselines. Inference GPU memory usage of the network evaluation step (Fig. \ref{fig:inference}) for different number of sampled locations.}
	\setlength\extrarowheight{0.605pt}
	\centering
	\label{tab:memoryresults}
	\begin{footnotesize}
	    \begin{tabular}{@{\extracolsep{4pt}}>{\raggedright\arraybackslash}p{0.125\columnwidth}>{\centering\arraybackslash}p{0.068\columnwidth}>{\centering\arraybackslash}p{0.115\columnwidth}>{\centering\arraybackslash}p{0.115\columnwidth}>{\centering\arraybackslash}p{0.068\columnwidth}>{\centering\arraybackslash}p{0.115\columnwidth}>{\centering\arraybackslash}p{0.115\columnwidth}}
	\hline
	\multirow{2}{0.19\columnwidth}{\vspace{-0.3cm}$\,$\\Model/\\OSS-Net\\Configuration} & \multicolumn{3}{c}{BraTS 2020}  & \multicolumn{3}{c}{LiTS} \\ 
	\cline{2-4} \cline{5-7}
	& Training $2^{14}$ loc. & Inference $2^{14}$ locations & Inference $2^{17}$ locations & Training $2^{14}$ loc. & Inference $2^{14}$ locations & Inference $2^{17}$ locations \\ 
	\hline
	3D res. U-Net & 14.41\si{\giga\byte} & \multicolumn{2}{c}{3.57\si{\giga\byte} (dense pred.)} & 14.41\si{\giga\byte}\textsuperscript{$\star$} & \multicolumn{2}{c}{3.57\si{\giga\byte} (dense pred.)\textsuperscript{$\dagger$}} \\
	2d res. U-Net & $\;\;\;$1.16\si{\giga\byte}\textsuperscript{$\ddagger$} & \multicolumn{2}{c}{0.46\si{\giga\byte} (slice pred.)\textsuperscript{$\ddagger$}} & $\;\;\;$4.29\si{\giga\byte}\textsuperscript{$\ddagger$} & \multicolumn{2}{c}{1.20\si{\giga\byte} (slice pred.)\textsuperscript{$\ddagger$}$\;\;$} \\
	O-Net \cite{mescheder2019} & $\;\;\;2.35\si{\giga\byte}$ & 0.29\si{\giga\byte} & 1.93\si{\giga\byte} & $\;\;\;5.07\si{\giga\byte}$ & 1.47\si{\giga\byte} & 1.99\si{\giga\byte} \\ 
	\hline 
	OSS-Net A & $\;\;\;2.58\si{\giga\byte}$ & 0.39\si{\giga\byte} & 2.73\si{\giga\byte} & $\;\;\;5.18\si{\giga\byte}$ & 1.50\si{\giga\byte} & 2.35\si{\giga\byte} \\ 
	OSS-Net B & $\;\;\;2.76\si{\giga\byte}$ & 0.48\si{\giga\byte} & 3.45\si{\giga\byte} & $\;\;\;5.21\si{\giga\byte}$ & 1.53\si{\giga\byte} & 2.53\si{\giga\byte} \\ 
	OSS-Net C & $\;\;\;2.59\si{\giga\byte}$ & 0.39\si{\giga\byte} & 2.73\si{\giga\byte} & $\;\;\;5.19\si{\giga\byte}$ & 1.51\si{\giga\byte} & 2.35\si{\giga\byte} \\ 
	OSS-Net D & $\;\;\;2.76\si{\giga\byte}$ & 0.48\si{\giga\byte} & 3.46\si{\giga\byte} & $\;\;\;5.22\si{\giga\byte}$ & 1.53\si{\giga\byte} & 2.53\si{\giga\byte} \\ 
	\hline
	\multicolumn{7}{l}{\textsuperscript{$\star$} Memory usage when trained on reduced resolution ($224^3$). Memory usage at native resolution $>110\si{\giga\byte}$.}\\
	\multicolumn{7}{l}{\textsuperscript{$\dagger$} Memory usage on downsampled resolution ($224^3$) with prediction upsampled to native res. ($512^3$).}\\
	\multicolumn{7}{l}{\textsuperscript{$\ddagger$} Memory usage for processing a single slice (native res.), memory increases linearly for multiple slices.}
\end{tabular}
	\end{footnotesize}
\end{table}
\indent We analysed the effect of the number of locations on the inference runtime (Fig. \ref{fig:runtime}), as the maximum number of locations evaluated per forward pass affects the memory consumption significantly (Tab. \ref{tab:memoryresults} \& Fig. \ref{fig:memory}). We also measured the runtime of the original O-Net inference approach in conjunction with OSS-Net variant C and compared this to the modified inference approach. Runtimes were averaged over the whole BraTS 2020 validation dataset, with an initial octree resolution of one-quarter of the BraTS 2020 resolution.\\
\indent The improved inference approach achieves a two times faster inference than the original approach, regardless of the maximum number of locations processed. The runtime for both inference approaches remained constant for $2^{12}$ and more maximum locations processed in a single forward pass (Fig. \ref{fig:runtime}). Taking both inference speed and memory consumption (Fig. \ref{fig:memory}) into account, an optimal maximum number of locations is in the range of $2^{12}$ to $2^{14}$. We evaluated the disagreement between voxel predictions for both inference approaches to be merely $5.9658\cdot 10^{-4}\%$ on the BraTS 2020 validation set. This demonstrates that the proposed inference approach retains segmentation performance, while begin two times faster.
	\section{Conclusion} \label{sec:conclusion}
We propose OSS-Net for memory-efficient and high-resolution semantic segmentation of 3D medical data in function space. OSS-Net employs local observations to overcome the limited expressiveness of the original occupancy networks, achieving segmentation performance comparable to state-of-the-art 3D CNN approaches. We experimentally demonstrated the subsequent significantly increased segmentation accuracy in comparison to the original O-Net, as well as in comparison to an efficiency baseline (slicing based approach, 2D residual U-Net). Compared to a voxelised CNN performance baseline (3D residual U-Net), OSS-Net performs on par in segmentation accuracy in the brain tumour segmentation on the BraTS 2020 dataset, while falling slightly short on a dataset with limited training data ($\sim 100$ liver segmentation samples, LiTS dataset). OSS-Net vastly outperforms the function space and efficient baselines on the latter more challenging dataset.\\
\indent In terms of training memory consumption (for a given resolution), OSS-Net is similarly efficient to the O-Net and in excess of 5 times more memory efficient than the voxelised CNN baseline (and 10 times more efficient at inference). In comparison to the efficient baseline, OSS-Net requires more memory at training (2.6GB in comparison to 1.2GB) and a similar amount during inference. Both memory requirements are feasible with mid-range consumer GPUs, and OSS-Net has the additional advantage of being able to capture 3D relations. The proposed OSS-Net inference approach, that leverages prior encoder predictions, improves the inference speed by a factor of two in comparison to the original O-Net. In summary, OSS-Net combines the strong segmentation performance of the voxelised CNN performance baseline with the memory efficiency of the original O-Net, enabling accurate, fast and memory efficient 3D semantic segmentation that can scale to high resolutions.\\
\indent Future work may consider a more principled approach to choosing the best sampling strategies, beyond the ablation study presented here. For example, using the network's confidence of past predictions could be employed to achieve an adaptive sampling strategy.
	
\noindent\textbf{Acknowledgements} We thank Marius Memmel and Nicolas Wagner for the insightful discussions, Alexander Christ and Tim Kircher for giving feedback on the first draft, and Markus Baier as well as Bastian Alt for aid with the computational setup.\\
This work was supported by the Landesoffensive f\"{u}r wissenschaftliche Exzellenz as part of the LOEWE Schwerpunkt CompuGene. H.K. acknowledges support from the European Research Council (ERC) with the  consolidator grant CONSYN (nr. 773196). O.C. is supported by the Alexander von Humboldt Foundation Philipp Schwartz Initiative.
	
	\bibliography{oss_net_bib}
	
\end{document}